\documentclass{article}

\newcommand{\uproman}[1]{\uppercase\expandafter{\romannumeral#1}}

\usepackage[english]{babel}
\usepackage{color}
\usepackage[a4paper,lmargin={2cm},rmargin={2cm},tmargin={2.5cm},bmargin = {2.5cm}]{geometry}
\usepackage{amssymb}
\usepackage{amsthm}
\usepackage{graphicx}
\usepackage{geometry}
\usepackage{amsmath}
\usepackage{xcolor}
\usepackage{listing}
\usepackage{wallpaper}
\usepackage{hyperref}
\usepackage{comment}
\usepackage[backend=biber, style= nature, sorting = none, maxbibnames=999]{biblatex}
\usepackage[font=footnotesize]{subcaption}
\usepackage{float}
\usepackage{stmaryrd}
\usepackage{setspace}
\usepackage{makecell}
\usepackage{acro}
\usepackage{array}
\usepackage{listings}
\definecolor{codegreen}{rgb}{0,0.6,0}
\definecolor{codegray}{rgb}{0.5,0.5,0.5}
\definecolor{codepurple}{rgb}{0.58,0,0.82}
\definecolor{backcolour}{rgb}{0.95,0.95,0.92}
\lstdefinestyle{mystyle}{
    backgroundcolor=\color{backcolour},   
    commentstyle=\color{codegreen},
    keywordstyle=\color{magenta},
    numberstyle=\tiny\color{codegray},
    stringstyle=\color{codepurple},
    basicstyle=\ttfamily\footnotesize,
    breakatwhitespace=false,         
    breaklines=true,                 
    captionpos=b,                    
    keepspaces=true,                 
    numbers=left,                    
    numbersep=5pt,                  
    showspaces=false,                
    showstringspaces=false,
    showtabs=false,                  
    tabsize=2
}
\lstdefinelanguage{Turtle}{
    morekeywords={@prefix, a},
    morestring=[b]",
}
\usepackage{subcaption}

\usepackage{enumitem}
\usepackage{tabto}

\usepackage{amsthm}
\usepackage{mathdots}

\usepackage{subcaption}

\title{An ontology-based description of nano computed tomography measurements in electronic laboratory notebooks: from metadata schema to first user experience}
\author{F. Kirchner$^{*,**,1}$, D.C.F. Wieland$^{*,1}$, S. Irvine$^{2}$, S. Schimek$^{1}$, J. Reimers$^{1,3}$, \\ R. Aversa$^{4}$, A. Boubnov$^{5}$, C. Lucas$^{6}$,  S. Flenner$^{2}$, I. Greving$^{2}$, A. Lopes Marinho$^{1}$,\\ T.M. Wong$^{1,2}$, R. Willumeit-Römer$^{1}$, C. Eschke$^{**,1}$, B. Zeller-Plumhoff$^{1,7}$}
\date{\small{*these authors have contributed equally to the manuscript}\\ \small{**corresponding authors: fabian.kirchner@hereon.de, catriona.eschke@hereon.de}
\\ \small{$^1$ Institute of Metallic Biomaterials, Helmholtz-Zentrum Hereon, Germany}
\\ \small{$^2$ Institute of Materials Physics, Helmholtz-Zentrum Hereon, Germany}
\\ \small{$^3$ Ernst Ruska-Centre for Microscopy and Spectroscopy with Electrons, Forschungszentrum Jülich GmbH, Jülich 52425, Germany} 
\\ \small{$^4$ Karlsruhe Institute of Technology, Scientific Computing Center (SCC), Kaiserstraße 12, 76131 Karlsruhe,
Germany}
\\ \small{$^5$ Karlsruhe Institute of Technology, Institute of Nanotechnology (INT), Kaiserstraße 12, 76131 Karlsruhe,
Germany}
\\ \small{$^6$ Bruker Daltonics SPR, Falkenried 88, 20251 Hamburg, Germany}
\\ \small{$^7$ Data-driven Analysis and Design of Materials, Faculty of Mechanical Engineering and Marine Technologies, University of Rostock, Germany}
}

\begin{document}
\maketitle

\section*{Abstract}

In recent years, the importance of well-documented metadata has been discussed increasingly in many research fields. Making all metadata generated during scientific research available in a findable, accessible, interoperable, and reusable (FAIR) manner remains a significant challenge for researchers across fields. Scientific communities are agreeing to achieve this by making all data available in a semantically annotated knowledge graph using semantic web technologies. Most current approaches do not gather metadata in a consistent and community-agreed standardized way, and there are insufficient tools to support the process of turning them into a knowledge graph. We present an example solution in which the creation of a schema and ontology are placed at the beginning of the scientific process which is then -- using the electronic laboratory notebook framework Herbie -- turned into a bespoke data collection platform to facilitate validation and semantic annotation of the metadata immediately during an experiment. Using the example of synchrotron radiation-based nano computed tomography measurements, we present a holistic approach which can capture the complex metadata of such research instruments in a flexible and straightforward manner. Different instrument setups of this beamline can be considered, allowing a user-friendly experience. We show how Herbie turns all semantic documents into an accessible user interface, where all data entered automatically fulfills all requirements of being FAIR, and present how data can be directly extracted \textit{via} competency questions without requiring familiarity with the fine-grained structure of the knowledge graph.

\section{Introduction}

When performing experimental measurements, specifically \textit{in situ} three-dimensional (3D) imaging, a large amount of research data is being generated. Firstly, the actual raw data is produced, which is of main interest to scientists. Secondly, metadata is produced, explaining the measurement's surroundings, such as information about how the devices were set up, or at what point different stages of the experiment took place.  It is clear that either part is not meaningful without the other, and it is general consensus in research communities, that all data should be sustainably stored in a FAIR way, \textit{i.e.}\,findable, accessible, interoperable, and reusable.\cite{wilkinson2016} Nevertheless, in particular metadata is typically stored in an unstructured way, often \textit{ad hoc} using spreadsheets or text documents, or inside a classical notebook. Furthermore, the terms used and level of detail of the recorded data may vary between different experiments.

Electronic laboratory notebooks (ELNs) offer the opportunity to store (meta-)data in a more structured manner and were broadly introduced and studied in the last decade~\cite{rubacha2011review, bird2013laboratory}. Higgins \textit{et al.} provided a comprehensive comparison recently~\cite{higgins2022considerations}, which highlighted that the median lifetime of ELN software packages was only 7 years.
The authors suggested that the development of ELNs along user needs, understanding of laboratory culture and ongoing commitment of institutional support were keys to the long-term successful adoption.
For material science research specifically, eCAT~\cite{goddard2009ecat} and eLabFTW~\cite{carp2017elabftw} were two of the most popular frameworks, which were both data-centric ELNs for resource management (\textit{i.e.}\,text, files and images, etc.).
However, as Kanza~\textit{et al.} investigated in a recent user study~\cite{kanza2017electronic}, semantic technologies, such as tagging, advanced semantic search, storing meta-data and link to ontologies, can help overcoming the adoption barrier of ELNs.
Later in its extended user study~\cite{kanza2019too}, the authors found that simple semantic technology such as tags searching of documents could not enhance scientists' performance due to the personalized behavior of annotation and searching among different areas of scientific research.
Hence, building the knowledge-centric framework using well-structured ontologies could be beneficial for ELNs, while there were only limited offers of ELNs using modern ontology technology for knowledge management in scientific research~\cite{kanza2019too}.
To the authors' best knowledge, the semantic electronic lab book \emph{Herbie} \cite{kirchner2024} is the first ontology-based ELN framework directly working on a knowledge graph. It was developed for interdisciplinary material science research in a laboratory environment based on scientific user experience.

In this work, a workflow to enable FAIR synchrotron radiation-based nano computed tomography (SRnCT) is presented. 
SRnCT is a 3D X-ray imaging technique used in particular to study material and biological systems at resolutions well below 100\,nm \cite{Claro2023_review,Zeller2021_Review}. 
\textit{In situ} SRnCT measurements are conducted to investigate material functionality and dynamic behaviour. 
Clearly, FAIR (meta)data management is pivotal if the data is to be used for follow-up analyses such as \textit{in silico} model validation. 
The workflow is designed as follows: Firstly, a specialized metadata schema for SRnCT experiments was created. This schema covers key metadata required to describe and reproduce the performed experiment. The procedure to establish the schema followed that of a metadata schema for scanning electron microscopy (SEM) \cite{joseph2021}. The SRnCT metadata schema was then transformed into an application level ontology, which was aligned with the mid-level \textbf{PR}ovenance \textbf{I}nformation for \textbf{M}Aterials science ontology (PRIMA) inside the research area \cite{PRIMA_page}. The actual collection of the metadata during the experiment was performed in the ELN Herbie. To this end, the ontology was extended by shapes constraint language (SHACL) documents. The SHACL documents were uploaded to Herbie, thus automatically creating user friendly web forms ready to be filled out by the performing scientists. This approach ensured that the metadata was captured, validated, and in particular FAIRly stored inside a semantically annotated resource description framework (RDF)-based knowledge graph. This data strategy is in line with well-established community best practices.\cite{jacobsen2020} We show how the data entries within the knowledge graph can be transformed into an XML document which adheres to the metadata schema using a SPARQL query. Furthermore the usefulness of the ontology is tested by creating SPARQL queries from scientifically relevant competency questions and running these against the knowledge graph.
The applicability of the presented approach to similar experimental setups is discussed.

\section{Methodology}
\subsection{Design of metadata schema}

The SRnCT schema was developed alongside a number of other schemas for characterization techniques. The first represented technique schema, for SEM, is well documented within reference \cite{joseph2021} and registered in MetaRepo~\cite{SEM_schema}, the metadata repository of the Nanoscience Foundries and Fine Analysis (NFFA)-EUROPE Pilot (NEP). Other techniques similarly represented in the form of metadata schemas include magnetic resonance imaging \cite{MRI_schema} and transmission electron microscopy \cite{TEM_schema}.

Whilst originally intended to serve as a schema for all X-ray nanoCT experiments, it was quickly established that the metadata requirements for measurements conducted at a synchrotron X-ray source are significantly more extensive, and somewhat separate to, measurements performed on a commercial laboratory-based tomography setup. One reason for this is that synchrotron beamlines are constructed according to specific demands with respect to their capabilities and the properties of the delivered X-ray beam from the synchrotron. This makes each beamline a unique device with respect to its configuration. For this reason, the proposed schema was then split into two. A lab CT schema \cite{labCT_schema} was created which may be more generally applied to microCT and nanoCT experiments alike, on commercial instruments within a laboratory setting. The SRnCT schema used as the basis for the project in this paper has been designed specifically for nanoCT experiments performed at a synchrotron source. Whilst it is intended to serve as a basis for other synchrotron beamlines, the schema has some additional details tailored for the specific nanotomographic endstation of the beamline P05 \cite{Flenner2020} at Petra III at Deutsches Elektronen Synchrotron (DESY, Hamburg), which is operated by Helmholtz-Zentrum Hereon. The schema has been initially implemented in the XML schema format. 

\begin{figure}
    \centering
    \includegraphics[width=0.5\linewidth]{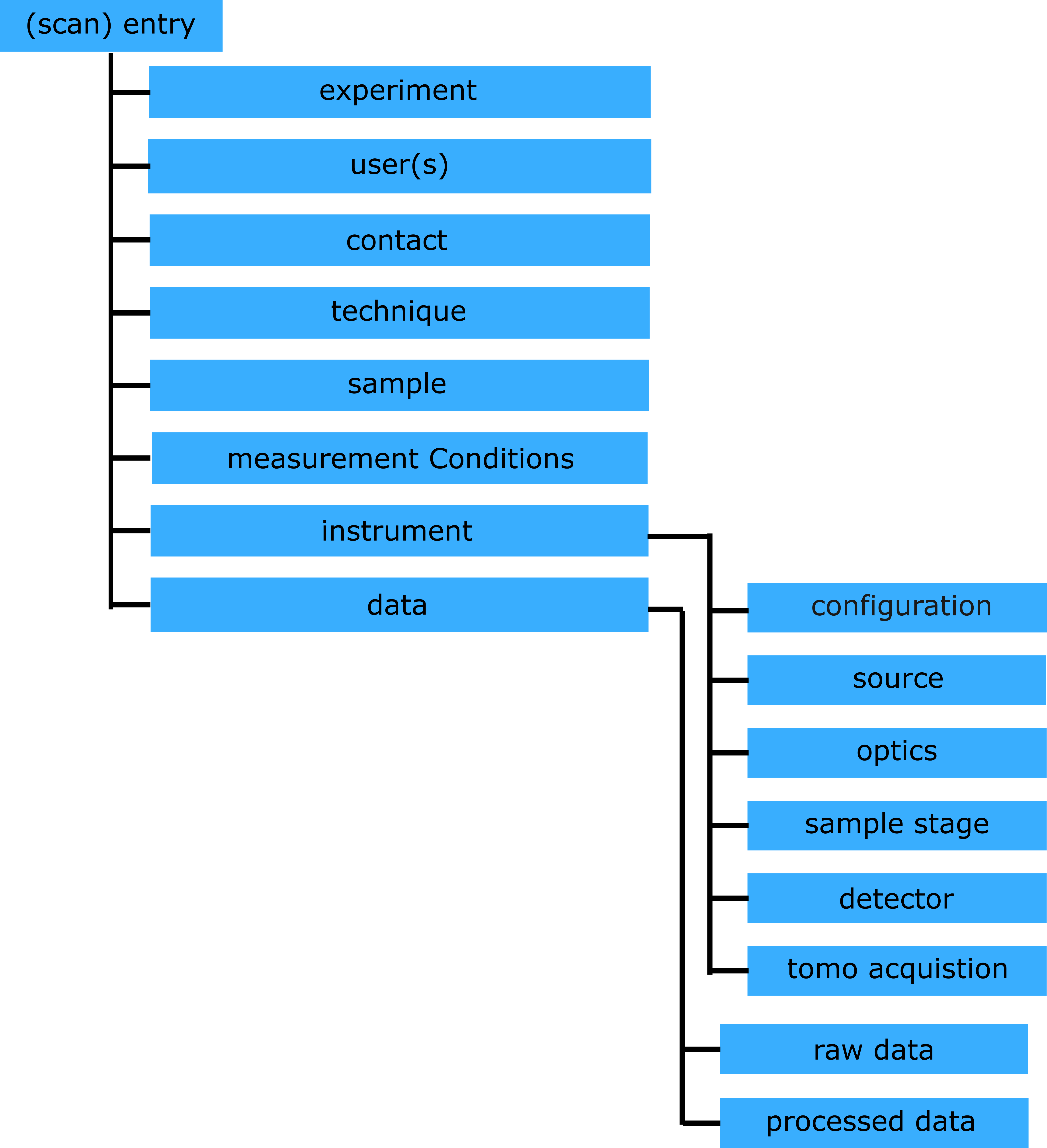}
    \caption{The hierarchy used in the schema to describe a scan measurement (entry).}
    \label{fig:schema_hierarchy}
\end{figure}

Following the general structure outlined in \cite{joseph2021}, the primary hierarchy of the template is outlined as shown in Figure~\ref{fig:schema_hierarchy}. Each information block is here referred to as a \textit{group}. The main groups are described below:

\textbf{Entry, or Scan Entry:} The entry level is the root element of the schema, resembling the NeXus \cite{Nexus2015}NXentry base class definition. It contains all the metadata describing a single measurement, \textit{i.e.}, a single SRnCT scan. 

\textbf{Experiment:} This group represents the overarching information pertaining to the overall experiment which may comprise multiple measurements. At a synchrotron this is generally referred to as the ‘beamtime’, which is generally linked to a beamtime proposal and assigned a unique identifier by the synchrotron facility. This identifier is also linked to the location of data storage. At many synchrotrons including Petra III, a JSON file containing the beamtime proposal data together with user data, is created before the beginning of the experiment period. The same information may be duplicated here for each scan entry within the same experiment.

\textbf{Contact:} The local contact is a designated person named from the beamline staff being responsible for helping in the beamtime organization before the experiment and assisting during the beamtime either by helping in performing the experiment or in case of problems. 

\textbf{Users:} This group represents the contact information of the user(s) responsible for the measurement, together with the indicated role of the user. The metadata properties were selected from the NeXus \cite{Nexus2015} NXuser group, adopting a similar naming convention. The Primary Investigator is usually the lead research user, while the Applicant is the person who submitted the beamtime proposal. It is typical for multiple users to be present during a beamtime to assist with measurements. 

\textbf{Technique:} This group states the measurement technique, which is in the case for the implemented beamline: TXM - absorption contrast,
TXM - Zernike phase contrast, TXM - inline phase contrast or Near Field Holotomography. 

\textbf{Sample:} This group describes the sample information, and can include any information describing the sample on which the measurement is performed, similar to the NeXus \cite{Nexus2015} NXsample group. This could be linked to the sample provenance by including a persistent uniform resource locator (PURL) if available.

\textbf{Measurement Conditions:} This group describes the conditions of the sample environment, which is used for \textit{in situ} testing, for example the usage of any special cells such as a flow cell, furnace or load frame. The environment itself is included together with the relevant physical parameters such as flow rate, temperature, or pressure.

\textbf{Instrument:} The instrument group describes the collection of the components of the beamline. Similar to the NXInstrument group, this group is modular whereby each component is by itself a group, and described below.

\textbf{Configuration:} This group contains information pertaining to the geometric setup of the beamline. For the nanotomography endstation at P05, two general configurations are possible, for the techniques of either Near Field Holography (NFH) or Transmission X-ray Microscopy (TXM) \cite{Flenner2020,Flenner2020_FZP,Flenner2022}. The current version of this template requires a selection of one of the two configurations/techniques. For the groups that follow, this selection then affects which parameters should be included for the Optics group such as geometric parameters including distances of the X-ray lenses or similar. 

\textbf{Source:} This group describes the source. The information included in this group may be split into electron source (and corresponding information regarding \textit{e.g.}\,storage ring current) details, and X-ray source or insertion device (\textit{e.g.}\,insertion gap) details.

\textbf{Optics:} This group describes the optical components used for the respective configuration, such as Fresnel zone plates. 

\textbf{SampleStage:} This group describes the sample stage motor positions.

\textbf{Detector:} This group describes the physical detector specifications.

\textbf{Tomo Acquisition:} Whilst related to the detector, this group specifically describes the key tomographic acquisition parameters. Also related to the raw \textbf{data} format.

\textbf{Data:} The data group describes the output datasets of the measurement. This includes information pertaining to the raw data acquired (data storage location, data format, etc) as well as any processed data which is generated after the necessary steps are performed such as phase retrieval and tomographic reconstruction.

\subsection{Implementation of schema in Herbie}

\subsubsection{Development of application ontology}

For semantic annotation of all metadata of the beamtime experiments, an application ontology was developed based on the PRIMA ontology~\cite{ahmad2025, PRIMA_page}. Like the metadata schema, this ontology is based on terms from the MDMC-NEP Glossary of Terms~\cite{aversa_2024}. The ontology was implemented in the web ontology language OWL 2.\cite{patelschneider2012}  Table~\ref{tab:prefixes} lists all prefixes used in the following. IRIs under \texttt{http://purls.helmholtz-metadaten.de/herbie/} will be registered with the Persistent Identifiers for Semantic Artifacts service (PIDA)~\cite{PIDA_page} provided by the Helmholtz Metadata Collaboration (HMC).

\begin{table}[]
    \centering
    \begin{tabular}{l|l}
         Prefix & IRI \\
         \hline
         \texttt{dash}              & \texttt{http://datashapes.org/dash\#} \\
         \texttt{foaf}              & \texttt{http://xmlns.com/foaf/0.1/} \\
         \texttt{hash}              & \texttt{http://purls.helmholtz-metadaten.de/herbie/hash/\#} \\
         \texttt{mbs}               & \texttt{http://purls.helmholtz-metadaten.de/herbie/mb/mbs/\#} \\
         \texttt{nfdi}              & \texttt{http://nfdi.fiz-karlsruhe.de/ontology/} \\
         \texttt{pmd}               & \texttt{https://w3id.org/pmd/co/} \\
         \texttt{prima}             & \texttt{https://purls.helmholtz-metadaten.de/prima/core\#} \\
         \texttt{prima\_experiment} & \texttt{https://purls.helmholtz-metadaten.de/prima/experiment\#} \\
         \texttt{prov}              & \texttt{http://www.w3.org/ns/prov\#} \\
         \texttt{qudt}              & \texttt{http://qudt.org/schema/qudt/} \\
         \texttt{rdfs}              & \texttt{http://www.w3.org/2000/01/rdf-schema\#} \\
         \texttt{sh}                & \texttt{http://www.w3.org/ns/shacl\#} \\
     \end{tabular}
    \caption{Namespace prefix bindings used in the text}
    \label{tab:prefixes}
\end{table}

For each mandatory block within the metadata schema, a matching subclass within the class hierarchy of PRIMA was added. For example, the ``entry'' block gives rise to the \texttt{mbs:ScanEntry} class which is a subclass of the \texttt{prima\_experiment:Measurement} class of the PRIMA ontology, or the ``instrument'' block is matched with the \texttt{mbs:BeamlineSetup} class, a transitive subclass of the \texttt{nfdi:Specification} class which is used in PRIMA as well. Similarly, nested blocks like ``tomo aquisition'' or ``detectors'' are mapped to the classes \texttt{mbs:TomoAcquisition} or \texttt{mbs:DetectorSetup}.

To enable structuring the resulting knowledge graph in line with PRIMA specification and to ensure high re-usability of collected data, content within one metadata block might be distributed among several classes.  For example, the ``instrument'' block specifies properties ``instrumentName'' and ``facilityName'', as well as properties like ``configuration'' and ``detector''.  The first two describe properties of the used beamline, which are a separate semantic concept from the ``instrument'' section covered by \texttt{mbs:BeamlineSetup}.  Therefore, a class \texttt{mbs:Beamline} (subclass of \texttt{prima:Instrument}) and a class \texttt{mbs:Facility} (subclass of \texttt{prima\_experiment:Laboratory}) were added.

Data properties such as ``pixelSize'' within the ``imagingDetails'' give rise to subclasses of \texttt{pmd:ValueObject}, in this case \texttt{mbs:ImagePixelSize}. PRIMA specifies that for each such datum the respective \texttt{pmd:ValueObject}-subclass is instantiated and linked to a \texttt{qudt:Quantity} instance, which references the actual numerical value alongside its unit.

\subsubsection{Implementation of SHACL shapes}

\begin{figure}
    \centering
    \includegraphics[width=1\textwidth]{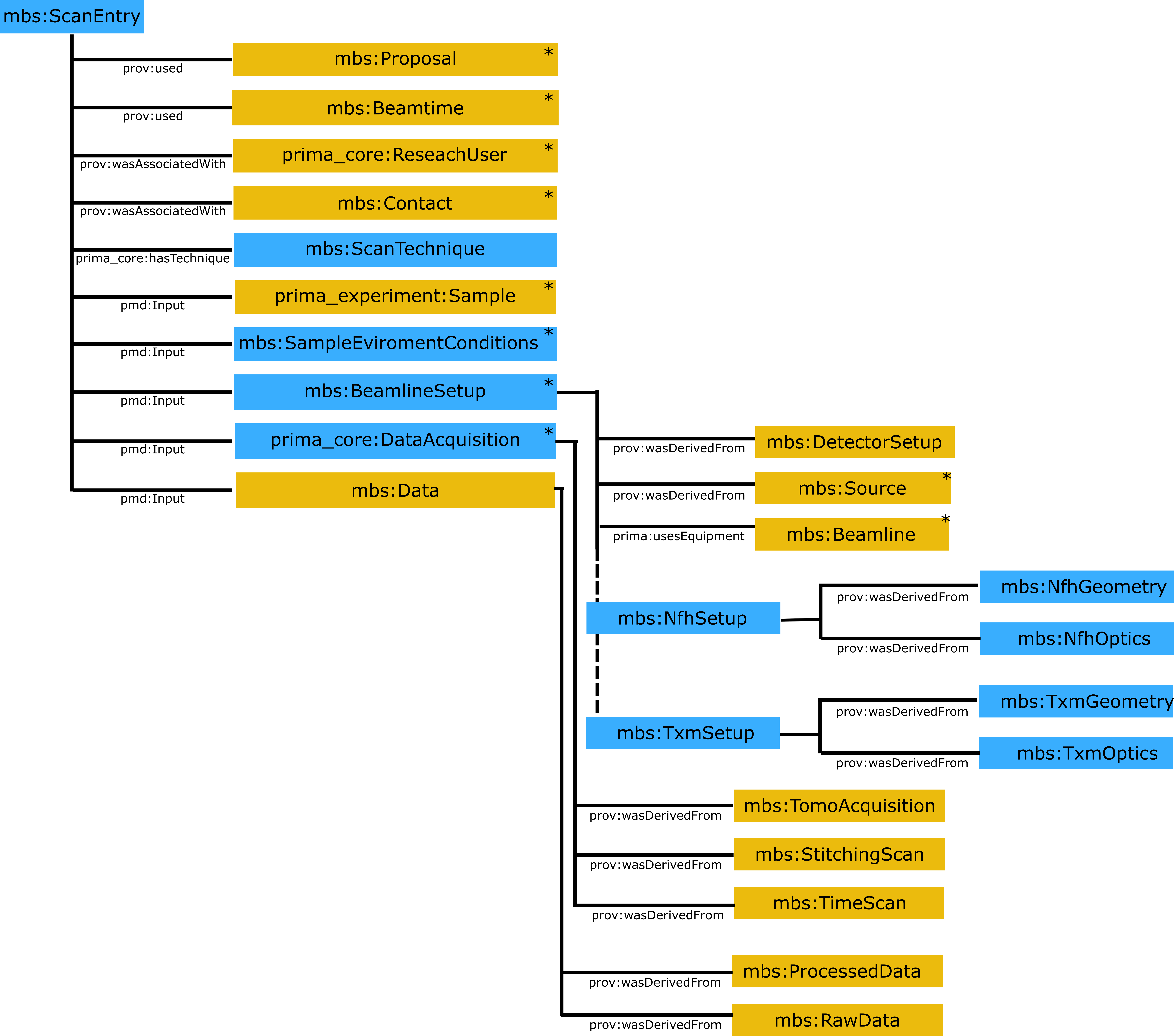}
    \caption{Hierarchy implemented in the SHACL-shapes to describe the experiment. The classes/shapes denoted by a star are single forms to be filled. Shapes marked in orange can most likely be reused in the development of a logbook for another beamline.}
    \label{fig:shacl_hierarchy}
\end{figure}

All metadata was recorded using the semantic lab notebook Herbie \cite{kirchner2024} whose user interface is configured by the ontology together with SHACL shapes \cite{knublauch2017}.  Herbie creates usable web forms for each node shape which specifies an \texttt{sh:targetClass}. Data entered \textit{via} such a form will be stored within an RDF knowledge graph. Additionally, these SHACL documents can also be used to validate externally provided knowledge graphs.

\begin{figure}
    \begin{subfigure}{0.48\textwidth}
        \centering
\begin{lstlisting}[language=Turtle]
:BeamlineShape a sh:NodeShape ;
  sh:targetClass mbs:Beamline ;
  hash:documentRoot true ;
  sh:property shared:rdfsLabel ;
  sh:property shared:rdfsComment ;
  sh:property :BeamlineShape_facility ;
.
  :BeamlineShape_facility sh:order 2 .

:BeamlineShape_facility a sh:PropertyShape ;
  sh:path prov:atLocation ;
  sh:qualifiedValueShape [
    sh:class mbs:Facility ] ;
  sh:qualifiedMinCount 0 ;
  sh:qualifiedMaxCount 1 ;
  dash:editor dash:InstancesSelectEditor ;
.
\end{lstlisting}
        \caption{Excerpt of SHACL implementation}
        \label{fig:shacl_documnet_beamline:code}
    \end{subfigure}
    \hspace*{\fill}
    \begin{subfigure}{0.48\textwidth}
        \includegraphics[scale=0.4]{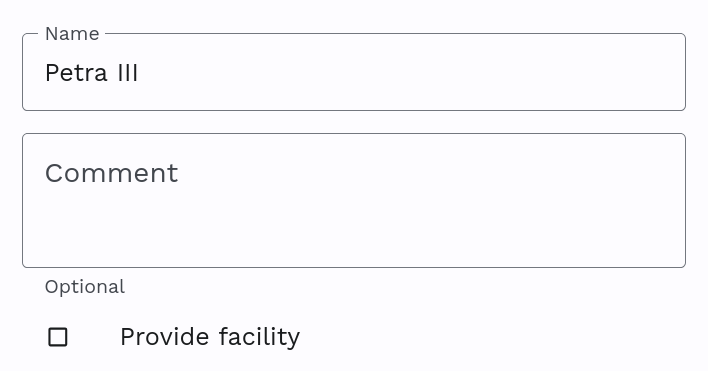}
        \caption{Resulting web form in Herbie}
    \end{subfigure}
    \caption{Example of a shape implementation and resulting user interface for generating instances of the \texttt{mbs:Beamline} class. In this example, a root node shape is generated specifying a \texttt{sh:targetClass}, e.g. \texttt{mbs:Beamline}, containing property shapes such as the facility name.}
    \label{fig:shacl_document_beamline}
\end{figure}

In order to structure the metadata collection process SHACL shapes were created following the strategy: All metadata blocks were segmented into sections, each of which could be created by the recording scientist. The general hierarchies and connection of the SHACL shapes are depicted in figure~\ref{fig:shacl_hierarchy}. For each section, a SHACL document was created containing one root SHACL node shape specifying an \texttt{sh:targetClass}, \textit{e.g.\,}the class \texttt{mbs:Beamline}, and containing property shapes for all required and optional parameters, such as its name or facility, as can be seen in figure~\ref{fig:shacl_document_beamline}. 

Nested metadata blocks were either inlined by nesting node shapes inside property shapes with \texttt{dash:editor dash:DetailsEditor}. These will be displayed as nested forms by Herbie.  Alternatively, only a property shape with \texttt{dash:editor dash:InstancesSelectEditor} and referencing the class via \texttt{sh:class} was added, if data within these blocks was to be reused among several experiments, as is the case for the \texttt{mbs:BeamlineSetup}. Such a property shape will be rendered by Herbie as a dropdown menu or a list of choice chips, depending on the number of selectable instances within the already existing knowledge graph.

\begin{figure}
    \begin{subfigure}{0.60\textwidth}
        \centering
\begin{lstlisting}[language=Turtle]
:DetectorSetupShape_ImagePixelSize a sh:PropertyShape ;
  sh:path prov:wasDerivedFrom ;
  sh:qualifiedValueShape [
    sh:class mbs:ImagePixelSize ;
    sh:property [
      sh:path qudt:quantityValue ;
      sh:node shared:Decimal_NanoM ;
      sh:minCount 1 ;
      sh:maxCount 1 ;
    ]
  ] ;
  sh:qualifiedMinCount 1 ;
  sh:qualifiedMaxCount 1 ;
.
\end{lstlisting}
        \caption{SHACL implementation via qualified property shape}
        \label{fig:shacl_document_image_pixel_size:code}
    \end{subfigure}
    \hspace*{\fill}
    \begin{subfigure}{0.38\textwidth}
        \centering
        \includegraphics[scale=0.4]{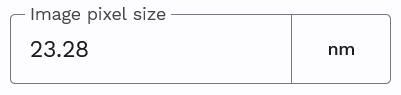}
        \caption{Resulting input field in Herbie}
        \label{fig:shacl_document_image_pixel_size:screenshot}
    \end{subfigure}
    \caption{Example of a shape implementation and resulting user interface on the basis of entering the image pixel size property.}
    \label{fig:shacl_document_image_pixel_size}
\end{figure}

Fields for the different parameters were created by including a property shape for the respective \texttt{pmd:ValueObject} subclass. Its \texttt{sh:maxCount} was set to \texttt{1} and the \texttt{sh:minCount} to \texttt{0} or \texttt{1}, depending on whether the parameter was optional or not. As the semantic distinction of these parameters was done \textit{via} subclassing \texttt{pmd:ValueObject}, the \texttt{sh:path} property of the property shape was usually set to the same property, in most cases \texttt{prov:wasDerivedFrom} as the parameter and its target are typically instances of some \texttt{prov:Entity} subclass. Therefore, the property shapes use \texttt{sh:qualifiedValueShape} to qualify all instances at the reused path, and \texttt{sh:qualifiedMinCount}/\texttt{sh:qualifiedMaxCount} are used instead of \texttt{sh:minCount}/\texttt{sh:maxCount}. At the \texttt{sh:qualifiedValueShape} property a node shape is embedded with \texttt{sh:class} set to the subclass of \texttt{pmd:ValueObject} corresponding to the parameter. Inside this node shape, the shapes for the actual numerical value and the unit are added.  To simplify the development, a set of reusable shapes was extracted, which contain property shapes for the \texttt{qudt:value} and \texttt{qudt:unit} properties. Despite this rather elaborate setup, each of these top-level property shapes is displayed by Herbie as a simple numerical text input with the respective unit.  Figure~\ref{fig:shacl_document_image_pixel_size} shows the implementation for the \texttt{mbs:ImagePixelSize}, as well as the resulting input field within Herbie. The implementation of the \texttt{shared:Decimal\_NanoM} shape was omitted for brevity.

\begin{figure}
    \begin{subfigure}{0.48\textwidth}
        \centering
\begin{lstlisting}[language=Turtle]
:BeamlineSetupShape a sh:NodeShape ;
  sh:targetClass mbs:BeamlineSetup ;
  # properties required in all beamline setups
  sh:or ( :NfhSetupShape :TxmSetupShape ) ;
.
:NfhSetupShape a sh:NodeShape ;
  sh:class mbs:NfhSetup ;
  # properties required for the NFH setup
.
:TxmSetupShape a sh:NodeShape ;
  sh:name "TXM"@en ;
  sh:class mbs:TxmSetup ;
  # properties required for the TXM setup
.
\end{lstlisting}
        \caption{SHACL implementation via \texttt{sh:or} construct}
        \label{fig:shacl_document_variants:code}
    \end{subfigure}
    \hspace*{\fill}
    \begin{subfigure}{0.48\textwidth}
        \includegraphics[scale=0.4]{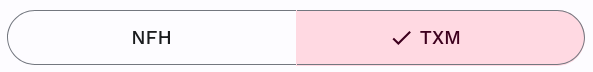}
        \caption{Resulting segmented buttons in Herbie}
        \label{fig:shacl_document_variants:screenshot}
    \end{subfigure}
    \caption{Example of a shape implementation and later web form for parameters depending on the setup of the beamline. One of two different configurations \texttt{mbs:NfhSetup} and \texttt{mbs:TxmSetup} can be selected.}
    \label{fig:shacl_document_variants}
\end{figure}

In cases where the set of required parameters varies depending on the class of experiments (configuration of the beamline), separate node shapes were created for each of these sets and then combined via \texttt{sh:or} in a top-level node shape. This was done for the \texttt{mbs:BeamlineSetup} which has subclasses \texttt{mbs:NfhSetup} and \texttt{mbs:TxmSetup} and whose instances require different parameters to be recorded.  Figure~\ref{fig:shacl_document_variants} shows the structure of the SHACL document and how in these cases Herbie renders a set of segmented buttons. Thus, the user can select one of the variants and is shown only those fields pertaining to the variant.

In cases where triples in the resulting data graph should be automatically generated, instances of \texttt{sh:SPARQLRule} were included.  For example, an \texttt{rdfs:label} was auto-generated for each \texttt{mbs:User} by concatenating their given and family name as well as their email-address. The concrete implementation can be seen in figure~\ref{fig:shacl_document_sparql_rule}.

\begin{figure}
    \centering
\begin{lstlisting}[language=Turtle]
:UserShape a sh:NodeShape ;
  sh:rule :UserShape_rdfsLabelRule ;
.
:UserShape_rdfsLabelRule a sh:SPARQLRule ;
  sh:condition [
    sh:property [ sh:path foaf:familyName ; sh:minCount 1 ] ;
    sh:property [ sh:path foaf:givenName ; sh:minCount 1 ] ;
    sh:property [ sh:path foaf:mbox ; sh:minCount 1 ] ;
  ] ;
  sh:prefixes foaf:, rdfs: ;
  sh:construct """
    CONSTRUCT { $this rdfs:label ?label }
    WHERE {
      $this foaf:familyName ?familyName .
      $this foaf:givenName ?givenName .
      $this foaf:mbox ?mbox .
      BIND(CONCAT(?givenName, " ", ?familyName, " (", ?mbox, ")") AS ?label)
    }
  """ ;
.
\end{lstlisting}
    \caption{SPARQL SHACL rule for automatically generating an \texttt{rdfs:label} from other user input}
    \label{fig:shacl_document_sparql_rule}
\end{figure}

To improve the general understandability of the generated web forms, the \texttt{sh:order} and \texttt{sh:group} features of SHACL were used to fix the ordering of the input fields as well as grouping them into sections.  Moreover,  the \texttt{sh:name} and \texttt{sh:description} properties were used to adjust the label and info text of the rendered input elements for property shapes covering paths from external ontologies, \textit{e.g.}\,\texttt{foaf:familyName} or \texttt{rdfs:label}.

\subsubsection{Connection to metadata schema}
Once all data for the beamtime experiment is collected inside a knowledge graph using Herbie, this data can be transformed into an XML document adhering to the metadata schema. This is done by querying the knowledge graph with a SPARQL construct query, which produces an RDF graph that contains all required data for the schema in a tree structure, and then serializing this RDF graph into an XML document which undergoes minor post-processing step removing RDFa-related tags. The SPARQL query is created in the following way:  Its CONSTRUCT part is a one-to-one resemblance of the tree structure of the XML schema with variables for each datum. The WHERE clause then maps every datum to the respective part in the knowledge graph. To ensure that only one tree is created for each \texttt{mbs:ScanEntry} instance, each node in the CONSTRUCT tree is given a unique IRI, either by binding an appropriate IRI in the knowledge graph, or by generating a new IRI from these.  See figure~\ref{fig:sparql_kg_to_xml} for an excerpt of the SPARQL query.

\begin{figure}
    \centering
\begin{lstlisting}[language=SPARQL]
CONSTRUCT {
  ?entry
    ns:entryID ?entryID ;
    ns:startTime ?startTime ;
  .
} WHERE {
  ?entry
    a mbs:ScanEntry ;
    dc:identifier ?entryID ;
    prov:startedAtTime ?startTime ;
  .
}
\end{lstlisting}
    \caption{SPARQL construct query for partially transforming Herbie's knowledge graph into a tree resembling the XML schema's structure}
    \label{fig:sparql_kg_to_xml}
\end{figure}

\subsection{Beamtime experiment}

The capabilities and versatility of the ELN was tested in an \textit{in situ} beamtime at the nanotomography endstation P05 at PETRA III (DESY, Hamburg, Germany). The aim was to investigate the degradation of a magnesium-based wires for biomedical applications under physiological conditions. Thus, the experiment was conducted using a custom bioreactor-coupled flow-cell setup optimized for \textit{in situ} SRnCT imaging. Detailed information on the experimental methodology and environment can be found in \cite{flowcel}. Initially, the selected magnesium (Mg)-based wire is immersed in a flow of ethanol (EtOH) for sterilization and one tomographic scan is obtained to capture the initial sample shape. Subsequently, the medium is changed to Dulbecco's modified Eagle's medium (DMEM) supplemented with 10\% fetal bovine serum (FBS) as degradation medium and multiple tomographic scans are recorded over time. For all measurements, a temperature of 37~°C is set, as well as a flow rate of 2~ml/min and a pH of 7.4.

This experiment was selected to evaluate the ELN's performance and usability, specifically examining the resulting entry for this experiment. The web browser Mozilla Firefox (Mozilla Corporation, San Francisco, USA, Version 132.0.2, 64-bit) was employed for accessing the Herbie web application. As the application resides on the internal network of the Helmholtz-Zentrum Hereon, secure access from the beamline was ensured \textit{via} a VPN connection established using GlobalProtect (Palo Alto Networks, Santa Clara, USA, Version 6.1.2-83). When accessing the platform, a dedicated workspace was created specifically for the beamtime, enabling systematic testing and application of the developed ontology.

\begin{figure}[htbp]
    \centering
    \includegraphics[width=0.8\textwidth]{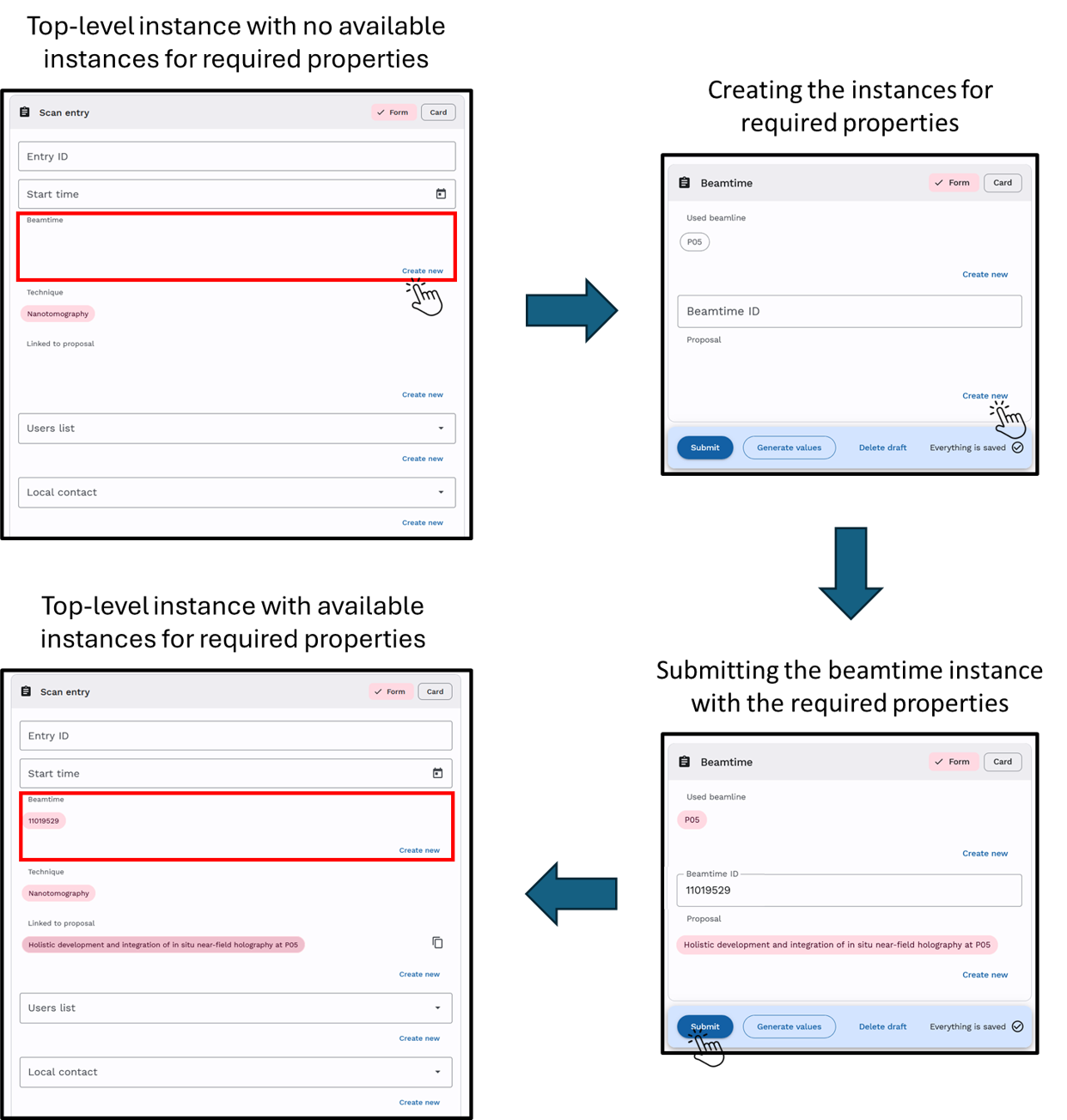}
    \caption{Schematic of the workflow using the ELN at the beamtime. Initially the the top level instance ``Scan entry'' had no required properties. By creating and submitting the missing instances, the necessary properties and instances can now be selected to complete the top-level instance.}
    \label{fig:workflow_herbie_beamtime}
\end{figure}

To record the measurement in the ELN, the web forms generated from the SHACL shapes were filled in, beginning with the top-level ``Scan entry'', as illustrated in Figure~\ref{fig:workflow_herbie_beamtime}. The hierarchical levels were sequentially completed. Missing instances were created in a step-by-step manner along with the respective sub-forms. Once the required resources were created, they appeared in the top-level entry form. As an example, in Figure~\ref{fig:workflow_herbie_beamtime}, the ``beamtime'' is initially a missing instance. By filling out this instance with its ``Beamtime ID'' and creating its sub-instance ``proposal'', the ``beamtime'' instance can be submitted. Since we have already created the ``proposal'' instance for the submission of the ``beamtime'', the proposal can also be selected in the top-level instance (Fig.~\ref{fig:workflow_herbie_beamtime}).\\
The generated instances and associated parameters are designed to remain permanent within the workspace. Thus, instances within classes such as ``beamtime'' remain unchanged for reuse in subsequent scans. This feature significantly reduces the workload for future experiments by minimizing repetitive data entry. However, if parameters change over the course of experiments or beamtimes, new instances and sub forms can be generated accordingly.

\subsection{Application of competency questions}

Finally, the developed application ontology was tested by translating each of the competency questions shown in table~\ref{tab:competency_questions} into a SPARQL select query and running this query against the knowledge graph which was generated during the beamtime experiment.  
\begin{table}[]
    \centering
    \begin{tabular}{rl}
           & Competency question \\
        \hline
         1 & What were the monochromator energy, flow rate, and system temperature during the measurement? \\
         2 & Which medium was used for the degradation? \\
         3 & What flow rate was applied? \\ 
         4 & Where is the raw data of the scan stored? \\
         5 & Which measurements were performed using a degradation cell and a bioreactor? \\
         6 & What FZP distance was used for the FZP ``QP040B.01''? \\
         7 & How often was the TXM or NFH technique applied? \\
         8 & What were the magnification and pixel size of the TXM or NFH experiments? \\
         9 & What was the ``sample\_in\_position''?
    \end{tabular}
    \caption{Competency questions for testing the ontology}
    \label{tab:competency_questions}
\end{table}

\section{Results and discussion}

\begin{figure}[htbp]
    \centering
    \includegraphics[height=0.9\textheight]{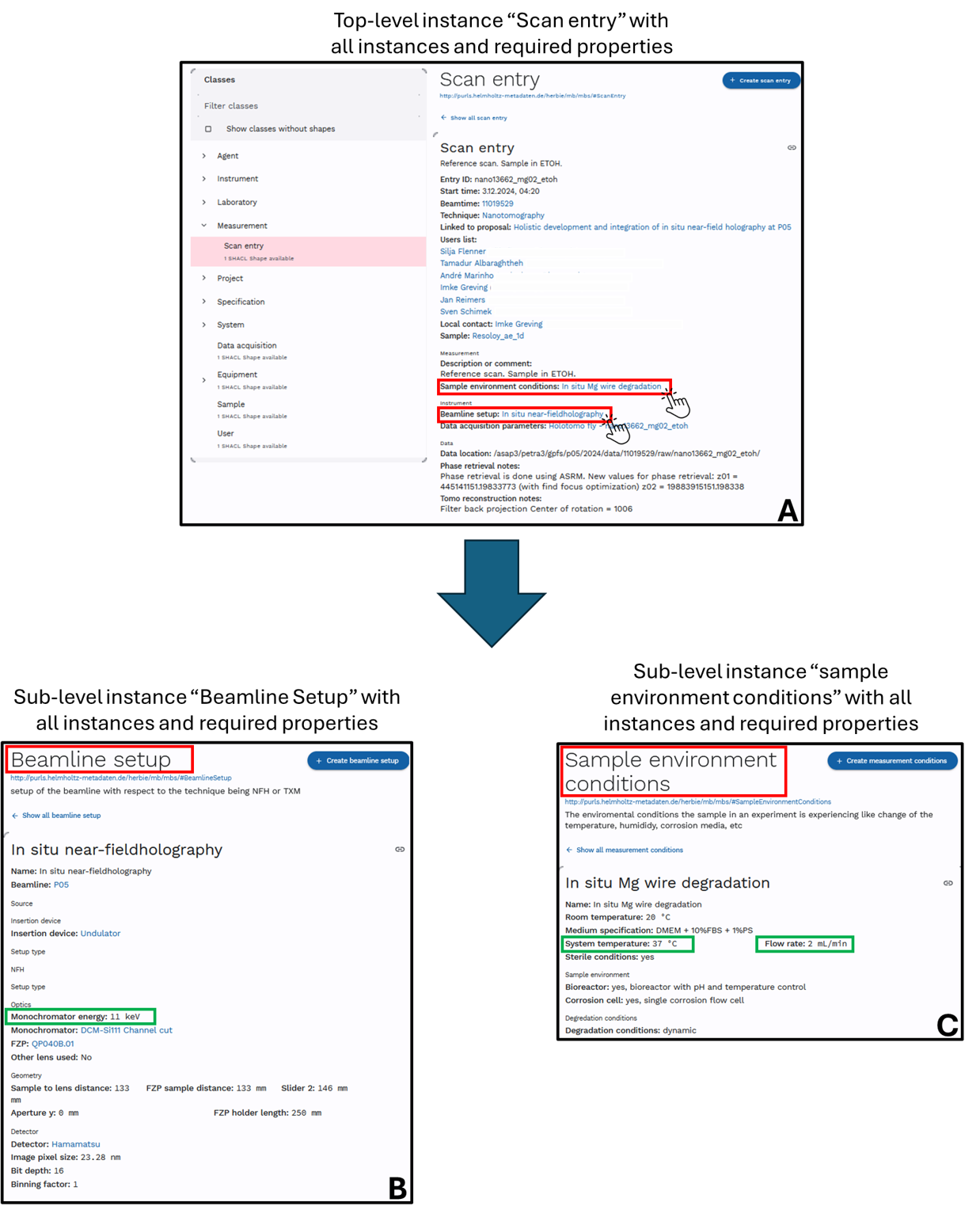}
    \caption{Final ELN entry for a nanoCT scan. A illustrates the top-level instance ``Scan entry'' with its sub-instances and the class selection menu on the left. B and C show the sub-level instances ``Beamline setup'' and ``Sample environment conditions'', respectively, with their required properties, which can be accessed via the hyperlink or the selection menu in A. Moreover, the addressed competency questions (c.f. Table~\ref{tab:competency_questions}) are highlighted in green boxes.}
    \label{fig:results_3_3_0000}
\end{figure}

A finalized ELN entry for a nanoCT scan of the Mg alloy wire immersed in EtOH is presented in Figure~\ref{fig:results_3_3_0000}. Figure~\ref{fig:results_3_3_0000}A illustrates the top-level instance ``Scan entry'' along with its subordinated instances and properties. This overview highlights the hierarchical organization of metadata fields and their corresponding sub-classes. By following the hierarchical structure, sub-level instances, such as the ``Beamline setup'' can be accessed for a more detailed examination of specific properties \textit{e.g.}\,``Image pixel size'', as shown in Figure~\ref{fig:results_3_3_0000}B. These sub-level instances can be inspected either by following the hyperlink in Figure~\ref{fig:results_3_3_0000}A or by searching for the instance in the classes overview menu on the left side of Figure~\ref{fig:results_3_3_0000}A. This allows for a comprehensive inspection of all sub-levels and properties.

During the testing process, the system exhibited stable performance with no major technical issues observed. The web-based interface functioned smoothly throughout. While the ELN proved to be effective for the experienced users that participated in development, further usability testing is necessary to assess its accessibility for a broader range of users, particularly those unfamiliar with the system.To enhance user-friendliness, the integration of automated selection for instrument-dependent parameters, such as ``x pixels'' which depend on the selected detector, can be generated automatically. This minimizes manual input and improves the system's adaptability to various experimental setups.

The evaluation and knowledge extraction of parameters is performed via competency questions that probe the knowledge graph.
Figure~\ref{fig:sparql_competency_question_1}, and Figures~\ref{fig:sparql_competency_question_4}, \ref{fig:sparql_competency_question_7}, and \ref{fig:sparql_competency_question_8} in Appendix~\ref{sec:sparql_appendix}, show the queries and results for some example competency questions. These questions were selected as they assess different parts of the experiment. The manner in which the data was structured within Herbie enables the extraction of deeply nested parameters without knowing the full tree structure, which simplifies later reuse. This nesting was performed with only a few properties, \textit{e.g.\,}\texttt{pmd:input} or \texttt{prov:wasDerivedFrom} and the semantic distinction of parameters was then achieved by having a subclass for each. So, for example in \ref{fig:sparql_competency_question_1}, the system temperature of the \texttt{mbs:ScanEntry} is retrieved by querying the property paths \texttt{pmd:input / prov:wasDerivedFrom*} for an instance of \texttt{mbs:SystemTemperature}. After the successful extraction of temperature and flow rate for example (question 1), further correlations can be established by connecting them to the results of the analysis and interpret differences between measurements. Similarly, the extracted parameters can be used for beamline alignment. A beamline setup is always slightly different due to the inherent complexity of such instruments, which,\textit{ e.g.}\,results in different motor positions of optical elements. Thus, competency questions like the question about the usage of the beamline, e.g. the magnification and pixel size (question 8), would enable the identification of settings where similar configurations were already used. Competency questions also allow for the extraction of comprehensive data from the ELN and might be subsequently used for analysis by machine learning. 

\begin{figure}
    \centering
\begin{lstlisting}[language=SPARQL]
SELECT ?energy ?energy_unit ?flow_rate ?flow_rate_unit ?temperature ?temperature_unit
WHERE {
  VALUES ?entry { <http://herbie.app.fzg.local/graph/scan-entrys/ZuUfmoGCbv58MzKJUYMK5B/> }

  ?entry a mbs:ScanEntry ;
    pmd:input / prov:wasDerivedFrom* [ a mbs:MonochromatorEnergy ;
      qudt:quantityValue [ 
        qudt:value ?energy ;
        qudt:unit / qudt:symbol ?energy_unit ] ] ;
    pmd:input / prov:wasDerivedFrom* [ a mbs:FlowRate ;
      qudt:quantityValue [
        qudt:value ?flow_rate ;
        qudt:unit / qudt:symbol ?flow_rate_unit ] ] ;
    pmd:input / prov:wasDerivedFrom* [ a mbs:SystemTemperature ;
      qudt:quantityValue [
        qudt:value ?temperature ;
        qudt:unit / qudt:symbol ?temperature_unit ] ] .
}
\end{lstlisting}
    \begin{tabular}{r|l|r|l|r|l}
         \texttt{energy} & \texttt{energy\_unit} & \texttt{flow\_rate} & \texttt{flow\_rate\_unit} & \texttt{temperature} & \texttt{temperature\_unit} \\
         \hline
         \texttt{11} & \texttt{"keV"} & \texttt{2} & \texttt{"ml/min"} & \texttt{37} & \texttt{"°C"}
    \end{tabular}
    \caption{SPARQL query and result to answer the competency question 1: What were the monochromator energy, flow rate, and system temperature during the measurement?}
    \label{fig:sparql_competency_question_1}
\end{figure}

The time needed to set up all semantic documents for the performed experiments highly depends on the familiarity of the scientists with the used semantic technologies RDF, OWL and SHACL. Similarly, if expert programmers were to implement the semantic documents, their understanding of the experimental workflow would be integral to the usability of the product. To provide a more objective overview of the required effort the number of classes, properties, node shapes and property shapes that had to be defined in order to facilitate the experiments can be quantified.

In the presented use case 147 new OWL classes were defined, see table~\ref{tab:classes} for a count with respect to their parent class. The majority were subclasses of \texttt{pmd:ValueObject} (71), \textit{i.e.}\,classes corresponding to individual numerical or textual data values. And although each of these requires some elaboration in the SHACL implementation, they would be highly reusable when modeling a related experimental setup where \textit{e.g.}\,used equipment has the same configuration parameters. Sub classes of \texttt{prima:Instrument} (12), \texttt{prima:System} (9), as well as \texttt{prima:Equipment} (7) are also good candidates for reuse.

\begin{table}[]
    \centering
    \begin{tabular}{l|r}
       \textbf{External class} & \textbf{No. of child classes} \\
       \hline
       \texttt{pmd:ValueObject}                                        & 71 \\
       \texttt{http://nfdi.fiz-karlsruhe.de/ontology/Specification}    & 20 \\
       \texttt{prima:Instrument}                                       & 12 \\
       \texttt{prima:Setting}                                          & 10 \\
       \texttt{prima:System}                                           & 9  \\
       \texttt{prima:Equipment}                                        & 7  \\
       \texttt{prima:DataAcquisition}                                  & 3  \\
       \texttt{qudt:QuantityValue}                                     & 2  \\
       \texttt{prima:Project}                                          & 2  \\
       \texttt{prima\_dataset:RawData}                                 & 2  \\
       \texttt{pmd:Object}                                             & 2  \\
       \texttt{prov:Agent}                                             & 1  \\
       \texttt{prima:Technique}                                        & 1  \\
       \texttt{prima\_dataset:Dataset}                                 & 1  \\
       \texttt{prima\_dataset:ProcessedData}                           & 1  \\
       \texttt{prima\_dataset:ReferenceData}                           & 1  \\
       \texttt{prima\_experiment:Laboratory}                           & 1  \\
       \texttt{prima\_experiment:Measurement}                          & 1
    \end{tabular}
    \caption{Number of classes in the application ontology for each external class from the mid- and top-level ontologies.}
    \label{tab:classes}
\end{table}

All SHACL documents contain a total of 172 node shapes of which 71 are actual named node shapes. Of these, 26 are node shapes specifying a \texttt{sh:targetClass}, and hence will be picked up by Herbie as an independent web form, see table~\ref{tab:node_shapes}. The majority of these root node shapes are for sub classes of \texttt{prima:Equipment} (7), \texttt{prima:Instrument} (4), and \texttt{prima:System} (4), and therefore reusable in a similar experimental setup. Finally, 207 property shapes were defined with a \texttt{sh:path} distributed as can be seen in table~\ref{tab:property_shapes}. 

The meta schemata which serves as the fundament to describe the information needed from an experiment differs from the realized structure within the ELN. By comparing figure~\ref{fig:schema_hierarchy} and figure~\ref{fig:shacl_hierarchy}, which show the structure of the hierarchy of the metadata schemata and SHACL shapes, it is obvious that both trees mostly overlap but have differences. These differences are a result of specific considerations during the implementation and adoption of special properties of the beamline P05 or usability. For instance, the information on the data acquisition is a child of the class ``Instrument'' in the metadata schemata and was moved to the top level in the ELN implementation. This decision was made as during a beamtime, the setup of the beamline usually does not change, whereas parameters on the scan, depending on the sample, might change. Due to this a user would be required only to create a new instance of \texttt{prima\_core:DataAcquistion} rather than generating a whole instance of \texttt{mbs:BeamlineSetup} again.  Nevertheless, each structure may be projected on the other, as shown above, and offers the degree of freedom needed for a successful and efficient integration. 

In this work, the presented metadata schema is a shorter template version, which can be extended to include many additional optional parameters, owing to the modular structure design. To adapt the used metadata setup to similar experiments, the ontology would require extension with the required classes and a corresponding extension of the set of SHACL documents would be necessary. 

The schema and implementation could for example be generalized to be applicable for synchrotron radiation-based microCT (SRµCT) measurements. Usually, SRµCT measurements require fewer optical components, thus, a smaller template may be sufficient.
To make the schema applicable for SRnCT measurements at different synchrotrons, further customized input may be required. Depending on the exact beamline layout different optics and other hardware components are used. Of course, a more generalized approach is feasible which would result in less customized entries and layout. However, this would also require less restricted entries and could lead to undefined string formatting of most entries. This would be a significant disadvantage for automated downstream analysis.

Depending on the exact overlap with the original setup a lot of code may be reused. Node shapes for the \texttt{qudt:Quantity}-instances can be directly reused if already present for the specific units or shapes for generating common properties to generate instances of \texttt{mbs:Proposal} or \texttt{mbs:Detector}. More specialized subclasses of classes already covered by a SHACL document which do not require any additional properties in the knowledge graph, can be included by adding the subclass to the ontology and a \texttt{sh:or} construct to the existing node shape. This then makes it possible for the user to select the more specific class in Herbie. An example is the \texttt{mbs:BeamlineSetup} which can be either an \texttt{mbs:NfhSetup} or an \texttt{mbs:TxmSetup}, both setups require similar parameters but also have their individual requirements. Figure~\ref{fig:shacl_document_variants} shows how they are combined in one SHACL document using \texttt{sh:or}. Figure~\ref{fig:shacl_hierarchy} also indicates those SHACL shapes with orange which can be reused immediately as they are for collecting the metadata at another beamline. Clearly, the major part of the code might be reused and expanding the already existing shape would consequently require significantly less work.

Extracting reusable parts of the schema into their own SHACL documents has the advantage that elements like specific \textit{e.g.}\,optical elements can be reused among several scans even from different beamtimes and have not to be entered a second time.  Also, this decreases the risk of different notations and increases reproducibility and findability within the ELN.  Additionally, if a separate SHACL document for the required class exists, Herbie renders a button next to the dropdown/choice chips to let the user conveniently create a instance with new parameters for the specific class.

When adapting or extending the setup,  the portions of the SPARQL construct query for transforming the knowledge graph into an XML document can be reused as well. The post-processing steps do not have to be adjusted. It is worth mentioning, that the SPARQL query -- like every data transformation pipeline -- potentially has to be adjusted if the XML schema, the ontology or the SHACL documents change.

In order to achieve FAIRness of meta data it is necessary to specify its semantics and validate its structure. Popular ELNs typically allow to immediately upload any not rigorously structured document. In these setups, usually, another framework is required to either extract well-structured and semantically annotated data or to run additional checks on the entered data, which validate conformance to some defined schema. So the workload of making all data FAIR is done in a post processing step. In our approach, we follow the concept of constructing an ontology and structuring the metadata up front before the ELN can be used. As a result, our approach leads to a very accessible ELN, which, out of the box, produces FAIR data, offers advanced semantic search capabilities, and automatically links all its data to ontologies.

\section{Conclusion and outlook}

The presented ELN forms are able to collect the main features of a nanoCT experiment at the nanotomography endstation of the beamline P05 and to collect all metadata needed to describe the experiments. Because of the structured and ontology-based approach, the metadata is completely semantically annotated.

In the future, different improvements are desired to ensure longterm use of the ELN forms. Currently, all data regarding the beamline information, such as motor positions, have to be filled in manually. However, many of the information are already logged in beamline specific files (either as ASCII file or as a NEXUS file) as they are required to describe the experiment. In future we are aiming to extract such information automatically so that the user needs to provide information on the sample and specific environments only. To do so, dedicated auxiliary scripts are needed to extract the information from log files depending on the file types and their structure. 

Due to the versatility of Herbie and the possibility to already adapt Herbie for the initial stage during \textit{e.g.}\,sample production, it is possible to comprehensively map the sample provenance. 
Additionally, future developments should focus on including the whole image processing workflow from phase retrieval, tomographic reconstruction, application of filters, segmentation and quantification within the ELN in an ontology-based manner. In doing so, a feedback loop between sample manufacturing and functionality can be established. Jalali \textit{et al.}\, showed that by further including large language models, the ELN provides great potential for material discovery.

As pointed out the parts of the SHACL shapes and ontology can be directly reused for other beamlines at the Petra III storage ring. Naturally, the ongoing development should be kept up to date with other metadata management efforts such as the DAPHNE4NFDI (DAta from PHoton and Neutron Experiments for NFDI) project\cite{barty_2023_8040606}. Importantly, interoperability between different implementations and ELN solutions is required and could be done using a tool like ELNdataBridge \cite{starman2024}, and an option for export/import of XML documents should be added. Moreover, mapping to data catalogues such as SciCat \cite{scicat} may be envisioned. Finally, to facilitate and accelerate SHACL shape creation, a graphical user interface may be introduced for this purpose.

\subsection*{Acknowledgements}

The SRnCT schema presented in this work was initially conceived as an additional technique represented within the collection of metadata schemas created by the MDMC-NEP Metadata Working Group. This is a collaboration between two projects: the Joint Lab ``Integrated Model and Data-Driven Materials Characterization'' (JL MDMC) and the Nanoscience Foundries and Fine Analysis (NFFA)-EUROPE Pilot (NEP). The authors thank Dr.~Reetu Joseph for her work in the creation of the metadata schema.

The authors acknowledge funding by the Helmholtz Association through the Joint Laboratory ``Model and Data-Driven Material Characterization'' (JL MDMC), the Helmholtz Imaging project SmartPhase (ZT-I-PF-4-027), the Helmholtz Metadata Cooperation projects MetaSurf (ZT-I-PF-3-076), and MetaCook (ZT-I-PF-3-046).
The authors acknowledge the ErUM-Data project ``KI4D4E: Ein KI-basiertes Framework für die Visualisierung und Auswertung der massiven Datenmengen der 4D-Tomographie für Endanwender von Beamlines'' which is funded by the Bundesministeriums für Bildung und Forschung (BMBF) (Förderkennzeichen 05D23CG1), and the EU’s H2020 framework program for research and innovation under grant agreement n. 101007417, NFFA-Europe Pilot.

\subsection*{Data availability}

All metadata collected during the experiment as well as the ontology, SHACL documents, and SPARQL queries of all competency questions will be made available upon reasonable request.

\printbibliography

\appendix
\section{Number of root node shapes and property shapes}
\label{sec:shape_appendix}
\begin{table}[H]
    \centering
    \begin{tabular}{l|r}
       \textbf{External class} & \textbf{No. of root node shapes} \\
       \hline
       \texttt{prima:Equipment}                                         & 7 \\
       \texttt{prima:Instrument}                                        & 4 \\
       \texttt{prima:System}                                            & 4 \\
       \texttt{http://nfdi.fiz-karlsruhe.de/ontology/Specification}     & 2 \\
       \texttt{prima:Project}                                           & 2 \\
       \texttt{prov:Agent}                                              & 1 \\
       \texttt{prima:DataAcquisition}                                   & 1 \\
       \texttt{prima:ResearchUser}                                      & 1 \\
       \texttt{prima\_experiment:Laboratory}                            & 1 \\
       \texttt{prima\_experiment:Measurement}                           & 1 \\
       \texttt{prima\_experiment:Sample}                                & 1
    \end{tabular}
    \caption{Number of root node shapes for a sub class of an external class from the mid- and top-level ontologies.}
    \label{tab:node_shapes}
\end{table}

\begin{table}[H]
    \centering
    \begin{tabular}{l|r}
      \textbf{Path} & \textbf{No. of property shapes} \\
      \hline
      \texttt{prov:wasDerivedFrom}     & 74 \\
      \texttt{qudt:quantityValue}      & 51 \\
      \texttt{prima:usesEquipment}     & 16 \\
      \texttt{prov:wasAttributedTo}    & 11 \\
      \texttt{pmd:characteristic}      & 7  \\
      \texttt{pmd:composedOf}          & 5  \\
      \texttt{rdfs:label}              & 4  \\
      \texttt{foaf:familyName}         & 4  \\
      \texttt{foaf:givenName}          & 4  \\
      \texttt{foaf:mbox}               & 4  \\
      \texttt{pmd:input}               & 4  \\
      \texttt{terms:identifier}        & 3  \\
      \texttt{terms:title}             & 2  \\
      \texttt{rdfs:comment}            & 2  \\
      \texttt{prov:used}               & 2  \\
      \texttt{prov:wasAssociatedWith}  & 2  \\
      \texttt{dcterms:identifier}      & 1  \\
      \texttt{prov:actedOnBehalfOf}    & 1  \\
      \texttt{prov:atLocation}         & 1  \\
      \texttt{prov:startedAtTime}      & 1  \\
      \texttt{prima:hasSetting}        & 1  \\
      \texttt{prima:hasTechnique}      & 1  \\
      \texttt{prima:performsAgentRole} & 1  \\
      \texttt{pmd:output}              & 1
    \end{tabular}
    \caption{Number of property shapes per RDF property.}
    \label{tab:property_shapes}
\end{table}

\section{Example SPARQL queries and results}
\label{sec:sparql_appendix}
Additional examples pertaining to selected competency questions from Table~\ref{tab:competency_questions}.
\begin{figure}
    \centering
\begin{lstlisting}[language=SPARQL]
SELECT ?file_location
WHERE {
  VALUES ?entry { <http://herbie.app.fzg.local/graph/scan-entrys/ZuUfmoGCbv58MzKJUYMK5B/> }

  ?entry a mbs:ScanEntry ;
    pmd:input / prov:wasDerivedFrom* [ a mbs:FileLocation ;
      rdfs:comment ?file_location ] .
}
\end{lstlisting}
    \begin{tabular}{l}
         \texttt{file\_location} \\
         \hline
         \texttt{"/asap3/petra3/gpfs/p05/2024/data/11019529/raw/nano13662\_mg02\_etoh/"}
    \end{tabular}
    \caption{SPARQL query and result to answer the competency question 4: Where is the raw data of the scan stored?}
    \label{fig:sparql_competency_question_4}
\end{figure}

\begin{figure}
    \centering
\begin{lstlisting}[language=SPARQL]
SELECT (COUNT(DISTINCT(?entry)) AS ?count)
WHERE {
  VALUES ?entry { <http://herbie.app.fzg.local/graph/scan-entrys/ZuUfmoGCbv58MzKJUYMK5B/> }
  VALUES ?setup { mbs:NfhSetup mbs:TxmSetup }

  ?entry a mbs:ScanEntry ;
    pmd:input [ a ?setup ] .
}
\end{lstlisting}
    \begin{tabular}{l}
         \texttt{count} \\
         \hline
         \texttt{1}
    \end{tabular}
    \caption{SPARQL query and result to answer the competency question 7: How often was the TXM or NFH technique applied?}
    \label{fig:sparql_competency_question_7}
\end{figure}

\begin{figure}
    \centering
\begin{lstlisting}[language=SPARQL]
SELECT ?parameter ?value ?unit
WHERE {
  VALUES ?entry {
    <http://herbie.app.fzg.local/graph/scan-entrys/ZuUfmoGCbv58MzKJUYMK5B/> }
  VALUES ?setup {
    mbs:NfhSetup
    mbs:TxmSetup }
  VALUES ?class {
    mbs:MagnificationFactorTheoretical
    mbs:MagnificationFactorMeasured
    mbs:ImagePixelSize }

  ?entry a mbs:ScanEntry ;
    pmd:input [ a ?setup ] ;
    pmd:input / prov:wasDerivedFrom* [ a ?class ;
      qudt:quantityValue ?quantityValue ] .

  ?class rdfs:label ?parameter .

  ?quantityValue qudt:value ?value .
  OPTIONAL { ?quantityValue qudt:unit / qudt:symbol ?unit } 
}
\end{lstlisting}
    \begin{tabular}{l|r|l}
         \texttt{parameter} & \texttt{value} & \texttt{unit} \\
         \hline
         \texttt{"theoretical magnification factor"@en} & \texttt{44.67} &               \\
         \texttt{"measured magnification factor"@en}    & \texttt{44.67} &               \\
         \texttt{"image pixel size"@en}                 & \texttt{23.28}  & \texttt{"nm"}
    \end{tabular}
    \caption{SPARQL query and result to answer the competency question 8: What were the magnification and pixel size of the TXM or NFH experiments?}
    \label{fig:sparql_competency_question_8}
\end{figure}

\end{document}